\documentclass[aps,prl,twocolumn,floats,showpacs]{revtex4}

\usepackage{graphicx}
\usepackage{amsmath}
\usepackage{amssymb}

\begin{document}
\title{Narrowband spectroscopy by all-optical correlation of broadband pulses}

\date{\today}
\author{Stanislav O. Konorov, Xiaoji G. Xu, John W. Hepburn and Valery Milner}

\affiliation{Department of Chemistry and The Laboratory for Advanced Imaging and Spectroscopic Research (LASIR), The University of British Columbia, Vancouver, Canada}

%-------------Abstract --------------------------------------
\begin{abstract}
High peak power ultrafast lasers are widely used in nonlinear spectroscopy but often limit its spectral resolution because of the broad frequency bandwidth of ultrashort laser pulses. Improving the resolution by achieving spectrally narrow excitation of, or emission from, the resonant medium by means of multi-photon interferences has been the focus of many recent developments in ultrafast spectroscopy. We demonstrate an alternative approach, in which high resolution is exercised by detecting narrow spectral correlations between broadband excitation and emission optical fields. All-optical correlation analysis, easily incorporated into the traditional spectroscopic setup, enables direct, robust and simultaneous detection of multiple narrow resonances with a single femtosecond pulse.

\end{abstract}

%82.53.Kp Coherent spectroscopy of atoms and molecules in physical chemistry and chemical physics)
%78.47.Fg Coherent nonlinear optical spectroscopy
%78.47.N- High resolution nonlinear optical spectroscopy
%78.47.+p Time-resolved optical spectroscopies and other ultrafast optical measurements in condensed matter
%42.50.Gy Effects of atomic coherence on propagation, absorption, and amplification of light; electromagnetically induced transparency and absorption
%42.65.Dr Stimulated Raman scattering; CARS
%33.20.Fb Raman and Rayleigh spectra (including optical scattering)
%33.20.Tp Vibrational analysis
%32.80.Qk Coherent control of atomic interactions with photons
%36.20.Ng Vibrational and rotational structure, infrared and Raman spectra
%39.30.+w Spectroscopic techniques

\maketitle

%----------Introduction 1: Resolution of ultrafast nonlinear spectroscopy -------------------
With recent advances of ultrafast laser sources, featuring high optical field amplitudes at low average power, coherent nonlinear spectroscopy with femtosecond (fs) pulses has been recognized as a promising tool for time-resolved chemical analysis \cite{Joo91,Schmitt97,Lang99}, remote sensing of complex molecules \cite{Scully02,Ooi05} and non-invasive imaging of biological objects \cite{Zumbusch99,Potma01,Cheng04}. Due to the broad spectral width of ultrashort pulses, femtosecond spectroscopy often lacks spectral resolution. The latter can be recovered by complementing the frequency resolved measurement with the time-delay scanning \cite{Heid01,Xu07}, or by combining the broadband fs excitation with the narrowband probing by picosecond pulses from a separate laser source \cite{Cheng02} or a spectral slice of a femtosecond source \cite{Oron03,Lim05} (``multiplex \textsc{cars}''). Alternatively, higher resolution can also be achieved without resorting to narrowband pulses and without the time-consuming delay scans. As many nonlinear optical methods involve interferences between multiple photons of different frequencies from under the same pulse bandwidth, improving the resolution is possible by controlling these interferences with spectral pulse shaping \cite{Weiner00}.  Spectral resolution, significantly better than the broad bandwidth of ultrashort laser pulses, has been achieved in second harmonic generation \cite{Zheng01}, multi-photon absorption \cite{Meshulach98,Pastirk03,Lozovoy03} and coherent Raman scattering \cite{Dudovich02,Oron02} of shaped femtosecond pulses.

%----------Introduction 2: CARS with fs pulses, the idea of resonance-induced correlations, NASCARS ----------
In this work we focus on Coherent Anti-Stokes Raman Scattering (\textsc{cars}) which has become a popular method in nonlinear optical spectroscopy and microscopy because of its high sensitivity to molecular structure \cite{Zheltikov00}. In \textsc{cars}, two laser photons, ``pump'' and ``Stokes'', excite the coherent molecular vibrations, which then scatter a third, ``probe'', photon generating the anti-Stokes signal. Despite the broad spectrum of the femtosecond pump and Stokes pulses, their collective two-photon field can be spectrally narrowed by means of the pulse chirping \cite{Nibbering92} or shaping \cite{Dudovich02}, providing selectivity of Raman excitation. On the other hand, shaping of the probe pulse have been used for narrowing the spectrum of the anti-Stokes emission \cite{Oron02}. In both cases, the spectral line narrowing is a result of a delicate interference of the input fields inside the Raman medium. As such, it is sensitive to the proper choice of the pulse shapes, non-resonant background and interferometrically stable time delays.

In the recently demonstrated method of noise-autocorrelation spectroscopy \cite{Xu08}, we have eliminated the requirement of an accurate pulse shaping, while achieving high spectral resolution without delay scanning and bandwidth multiplexing. The method is based on the detection of optical field correlations in the coherently scattered light. The correlations are induced by the vibrational resonances, and therefore carry the information about the Raman spectrum of the medium. The auto-correlation of the Raman spectrum has been extracted via computer post-processing of the measured data, which involved numerical correlation analysis with averaging over multiple noise realizations.

%-----------In this work...: Novelty of the method----------------------------------------------------------
Here, we show that spectral correlations in the coherently scattered broadband anti-Stokes radiation can produce a narrowband optical Raman spectrum similar to that obtained with spontaneous Raman or multiplex \textsc{cars}, yet without the narrowing of the pulse bandwidth. We take advantage of the possibility to detect optical correlations between two broadband pulses by means of the non-resonant interferences in the sum-frequency generation (\textsc{sfg}) of those pulses in the external nonlinear crystal \cite{DielsRudolphBook}. We start with the two pulses (hereafter called ``conjugate'' probe, $E_{pr}$, and reference, $E_{ref}$) designed to exhibit narrow single-peak \textsc{sfg} spectrum,
\begin{equation}\label{E_pr_ref}
E_{pr+ref}(\Omega )\propto \left|\int E_{pr}(\omega ) E_{ref}(\Omega -\omega ) d\omega\right|^2 \propto \delta (\Omega -\omega_{c}),
\end{equation}
where $\delta $ is the Dirac $\delta $-function and $\omega _{c}$ represents the central frequency of the probe/reference \textsc{sfg} spectrum. The probe pulse is then scattered off the molecular vibrations, generating the anti-Stokes pulse. Finally, the resonance-induced correlations in the anti-Stokes light are optically processed by mixing the \textsc{cars} signal with the reference pulse. The measured \textsc{cars}/reference \textsc{sfg} spectrum, $I_{\textsc{cars}+ref}(\Omega )\propto \left|\int E_{\textsc{cars}}(\omega ) E_{ref}(\Omega -\omega ) d\omega\right|^2 $, reflects the narrow vibrational modes of the medium as described below.

%--------Theoretical part 1: spectral narrowing -----------------------
\begin{figure}
\centering
\includegraphics[width=0.98\columnwidth]{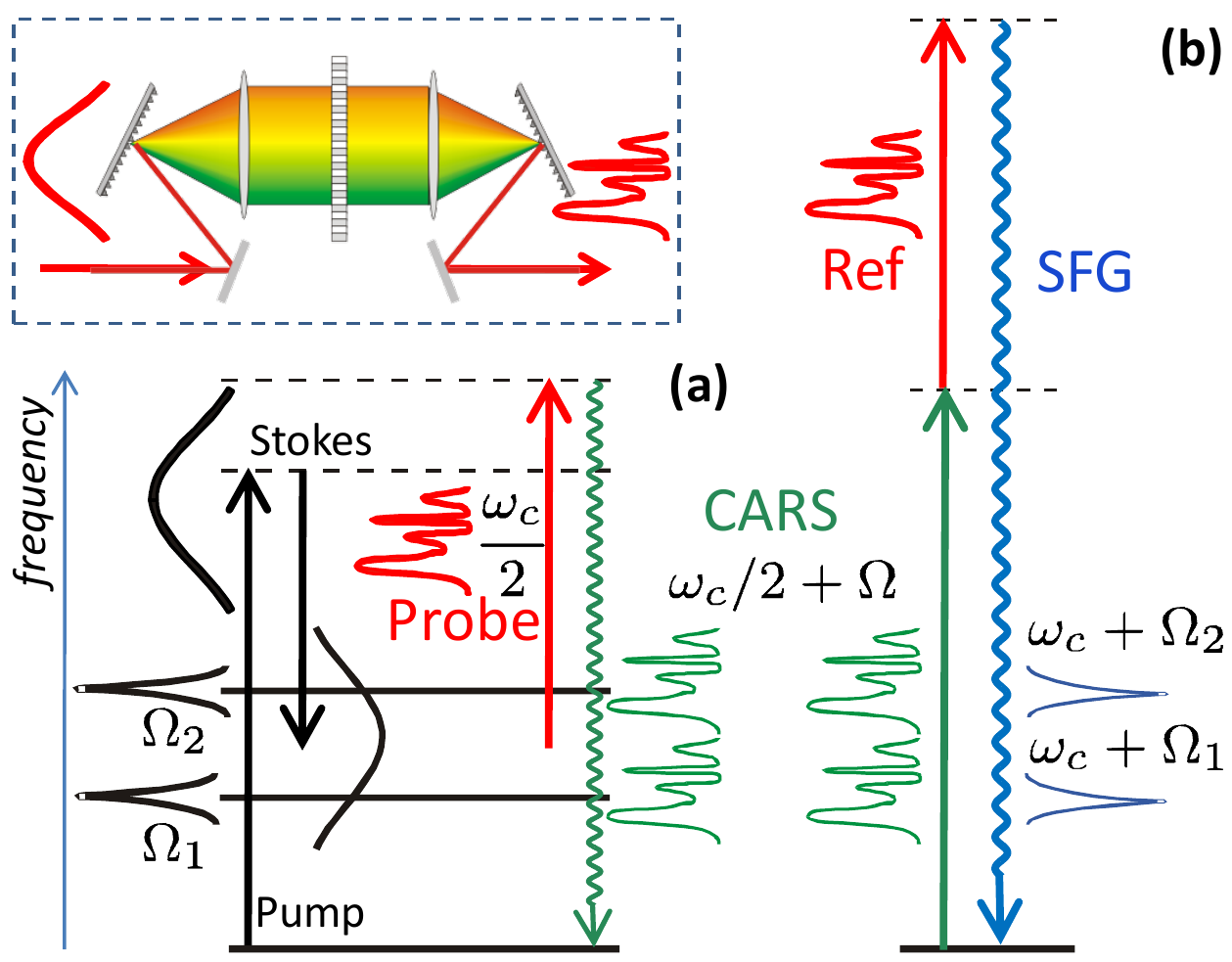}
\caption{ (Color online) (a) Schematic representation of the coherent anti-Stokes scattering of the spectrally shaped probe pulse (red) off the molecular vibrations at $\Omega _{1}$ and $\Omega _{2}$, induced by the transform-limited pump and Stokes pulses (black). (b) an all-optical retrieval of the narrowband Raman spectrum by means of the sum-frequency mixing of \textsc{cars} and reference pulses in a separate nonlinear crystal. In this work, reference pulse is a replica of the probe pulse, spectrally shaped by means of the liquid crystal based pulse shaper (inset).}
  %\vskip -.1truein
  \label{FigCARSscheme}
\end{figure}

In \textsc{cars}, coherent molecular vibrations are excited by a two-photon field of the simultaneously applied transform-limited pump and Stokes pulses, $E_p$ and $E_S$, respectively (black lines in Fig.\ref{FigCARSscheme}(a)). The induced vibrational coherence can be expressed as a product of the excitation spectrum and the spectrum of the Raman modes \cite{Xu07}:
\begin{equation}\label{R_of_Omega}
    R(\Omega )= A(\Omega ) \times \left(\sum\nolimits_k C_k(\Omega )+C_{\textsc{nr}}\right),
\end{equation}
where $A(\Omega )=\int E_p(\omega )E_S(\Omega -\omega )d\omega $, $C_k(\Omega )$ represents the contribution of the $n$th resonance, and the non-resonant background is denoted $C_{\textsc{nr}}$. Anti-Stokes scattering of the probe field, $E_{pr}$, off the molecular vibrations results in the \textsc{cars} signal, $E_{\textsc{cars}}$, which in the time domain can be written as:
\begin{equation}\label{E_CARS_of_t}
    E_{\textsc{cars}}(t) \propto R(t) E_{pr}(t),
\end{equation}
with $R(t)$ being the Fourier transform of the spectral response, $R(\Omega )$. The sum-frequency generation field of the \textsc{cars} and reference pulses is, therefore:
\begin{equation}\label{E_SFG_of_t}
    E_{\textsc{cars}+ref}(t) \propto E_{\textsc{cars}}(t) E_{ref}(t) = E_{pr}(t) E_{ref}(t) R(t),
\end{equation}
which in the frequency domain corresponds to the convolution of the probe/reference \textsc{sfg} spectrum, $E_{pr+ref}(\omega )$, and the vibrational spectrum of the medium:
\begin{equation}\label{E_SFG_of_omega}
    E_{\textsc{cars}+ref}(\omega_c+\Omega)= \int E_{pr+ref}(\omega_c-\Delta ) R(\Omega+\Delta ) d\Delta .
\end{equation}
Thus, by narrowing the spectrum of $E_{pr+ref}$ (Eq.\ref{E_pr_ref}), one can establish an exact correspondence between $E_{\textsc{cars}+ref}(\omega _{c}+\Omega )$ and $R(\Omega )$, retrieving the Raman frequency shifts directly from the measured \textsc{sfg} signal.

%-----------Theoretical part 2: self-conjugate probe/reference----------------------------
Unlike the all-optical processing used in multiplex \textsc{cars} and based on the resonant interference between the excited vibrational states and their precise amplitude and phase shaping \cite{Oron04}, our approach does not require any prior knowledge on the vibrational frequencies of the molecules. It is also not sensitive to the particular spectral shape of the probe pulses as long as the conjugation condition of Eq.\ref{E_pr_ref} is satisfied. The conjugate probe and reference pulses can be generated by various means, including simple frequency chirping or frequency down conversion \cite{Dayan04}. Here, we employ spectral pulse shaping to create a ``self-conjugate'' pulse, i.e. a broadband pulse with a narrowband second harmonic, $E_{pr+ref}(\Omega ) \rightarrow E_{pr+pr}(\Omega )$. The self-conjugate field serves both as the probe and the reference, thus eliminating the need of dealing with two separate pulses and simplifying the experimental setup. This result, shown schematically in Fig.\ref{FigCARSscheme}(b), provides the basis for the experimental method demonstrated in this work.

Self-conjugate pulses can be generated by shaping the original transform-limited femtosecond pulse with a pseudo-random binary phase mask \cite{Comstock04}. As a two-photon process, second harmonic generation (\textsc{shg}) at any given frequency $\omega _c$ is defined by the interference of multiple photon pairs, whose total frequency adds up to $\omega _c$. For this to result in a narrowband spectrum, the interference must be constructive at $\omega _c$ and destructive everywhere else. The first requirement is known to be satisfied by the condition $E_{pr}(\omega_c/2-\Delta) = E^*_{pr}(\omega_c/2+\Delta)$ \cite{Meshulach98}, which can be realized by applying an anti-symmetric spectral phase shaping around $\omega_c/2$. Destructive interference can be achieved by randomizing the spectral phase of the pulse while preserving the anti-symmetric property \cite{Zheng00,Comstock04}.

%------------Theoretical part 3: numerical simulations and optimization ------------------
\begin{figure}
\centering
\includegraphics[width=0.98\columnwidth]{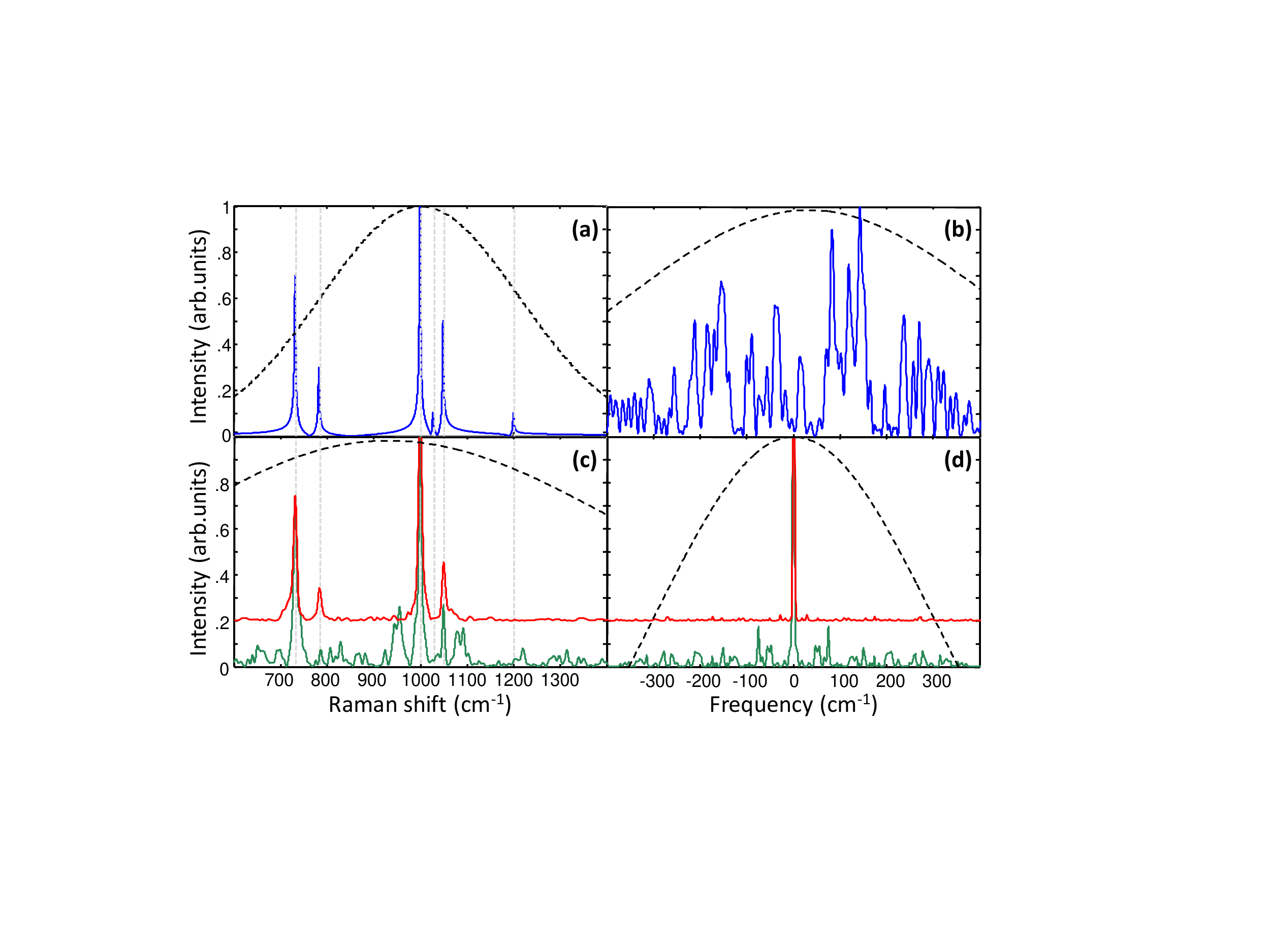}
\caption{(Color online) Numerical simulations of the proposed method. (a) Model spectrum of a toluene/orto-xylene mixture. (b) Calculated \textsc{cars} spectrum for the case of a pseudo-randomly shaped probe (see text). (c,d) Calculated spectra of the \textsc{cars}/probe \textsc{sfg} and probe \textsc{shg} signals, respectively, for the case of the non-optimized (green) and optimized (red) pseudo-random probe shaping. The line width is defined by the frequency resolution of the model pulse shaper (4.7 cm$^{-1}$), rather than by the overall bandwidth of the probe pulse (dashed lines).}
  %\vskip -.1truein
  \label{FigSimulations}
\end{figure}

Calculations of the sum-frequency mixing of the probe and \textsc{cars} pulses (Eq.\ref{E_SFG_of_omega}), and its correspondence to the model vibrational spectrum of the medium, are shown in Fig.\ref{FigSimulations}(a,c). First, an uncorrelated white noise, anti-symmetrized around the central frequency, has been added to the phase of the originally transform-limited probe field through a model pulse shaper. Owing to the presence of Raman resonances and despite the uncorrelated noise in the probe, spectral correlations are created in the \textsc{cars} field. Though not apparent in the \textsc{cars} signal itself (Fig.\ref{FigSimulations}(b)), they produce sharp peaks in the \textsc{cars}/probe \textsc{sfg} spectrum (Fig.\ref{FigSimulations}(c)), which can be clearly assigned to the corresponding Raman lines.

Unfortunately, residual correlations in the random phase give rise to the artificial lines and lower signal-to-noise ratio, seen both in the \textsc{cars}/probe \textsc{sfg} and probe/probe \textsc{shg} spectra (Fig.\ref{FigSimulations}(c) and (d),respectively). To improve the quality of the signal, we performed a numerical optimization based on the genetic algorithm search of a pseudo-random binary phase mask (constrained to $0$ and $\pi $ radian only \cite{Comstock04}) resulting in the narrowest calculated probe \textsc{shg} spectrum. Due to the finite spectral resolution of the pulse shaper and, therefore, finite width of the optimized \textsc{shg}, the correspondence between $E_{\textsc{cars}+pr}(\omega_c +\Omega )$ and $R(\Omega )$ in Eq.\ref{E_SFG_of_omega} is not exact. Hence, using the \textsc{cars}/probe \textsc{sfg} line width as a fitness parameter improved the performance of the method even further, but required \textit{a priori} assumption about the line width of the Raman resonances. The resulting pseudo-random binary masks were used in the experiments described below.

%---------Experimental part 1: Setup--------------------------------------------
\begin{figure}
\centering
\includegraphics[width=0.98\columnwidth]{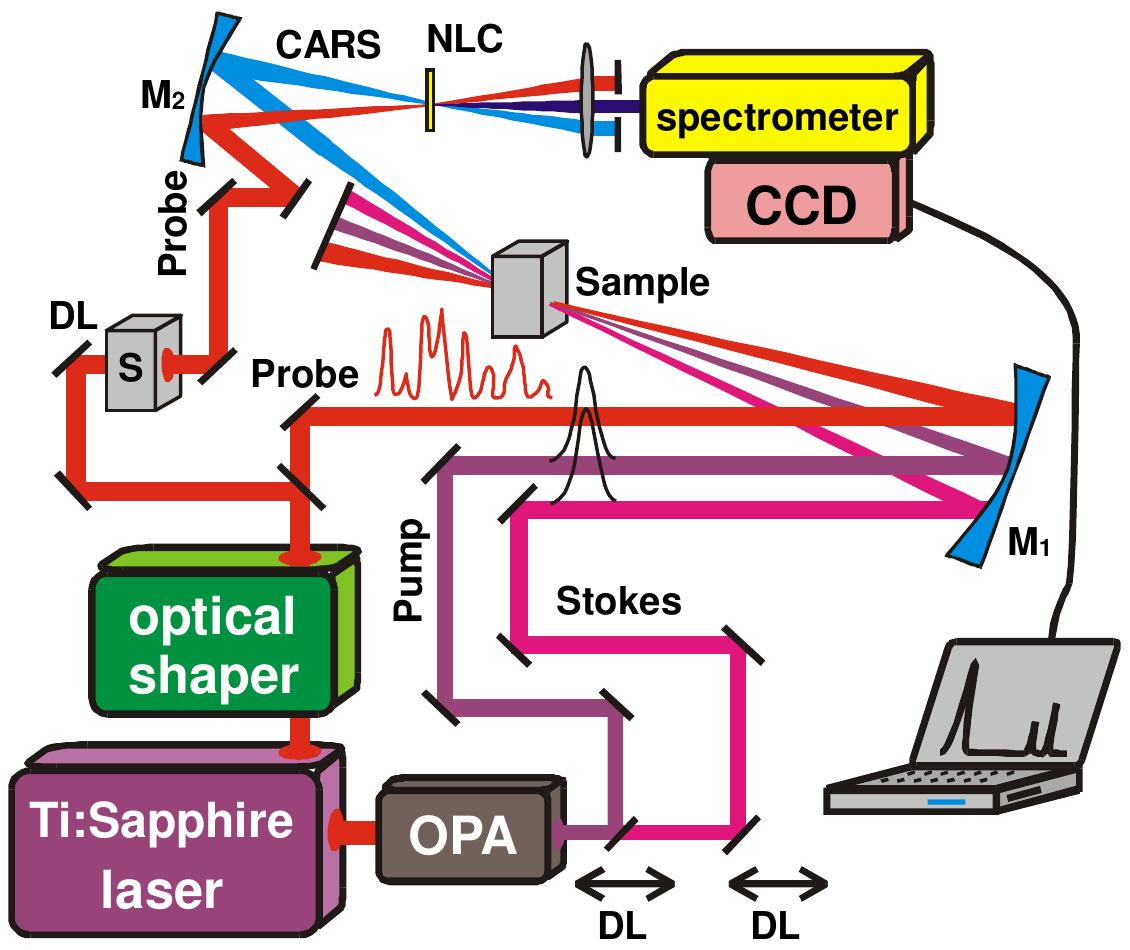}
\caption{Experimental setup. Transform-limited pump and Stokes pulses from an optical parametric amplifier (OPA) and a shaped probe pulse from the pulse shaper are focused on a 5 mm thick sample by a 20 cm focal length mirror (M$_{1}$). Output \textsc{cars} pulse is then focused by a similar mirror (M$_{2}$) on a nonlinear crystal (NLC), together with a replica of the probe pulse, to produce a sum-frequency generation signal, which is detected by a spectrometer (spectral resolution of 0.05 nm). The probe beam, used as a reference, is passed through an identical sample cuvette (S) to compensate the accumulated dispersion of the \textsc{cars} signal. DL - delay lines.}
  %\vskip -.1truein
  \label{FigSetup}
\end{figure}
The experimental setup (Fig.\ref{FigSetup}) was based on a standard Ti:Sapphire regenerative amplifier, producing 40 fs probe pulses at the central wavelength of 800 nm and 1 KHz repetition rate. An optical parametric amplifier generated the pump and Stokes beams at 1111 and 1250 nm, respectively. Pre-calculated spectral phase has been applied to the probe pulses by means of a home-made pulse shaper based on a liquid-crystal spatial light modulator (see inset in Fig.\ref{FigCARSscheme}) and provided spectral resolution of 10 cm$^{-1}$. Pump, Stokes and probe pulses, about 2 $\mu $J each, have been focused by a silver coated mirror into a quartz cuvette with liquid toluene, where they overlapped in time and space in the standard \textsc{boxcars} phase-matching geometry. \textsc{cars} signal has been then collimated and focused onto a 100 $\mu $m BBO crystal, together with the reference pulse which has been split off the shaped probe pulse in front of the sample. Sum-frequency signal from the nonlinear mixing of the \textsc{cars} and reference pulses has been coupled into a high-resolution spectrometer equipped with a CCD camera, and typically accumulated for about 1 minute.

%------------Experimental part 2: Results and discussions----------------------------
\begin{figure}
\centering
\includegraphics[width=0.98\columnwidth]{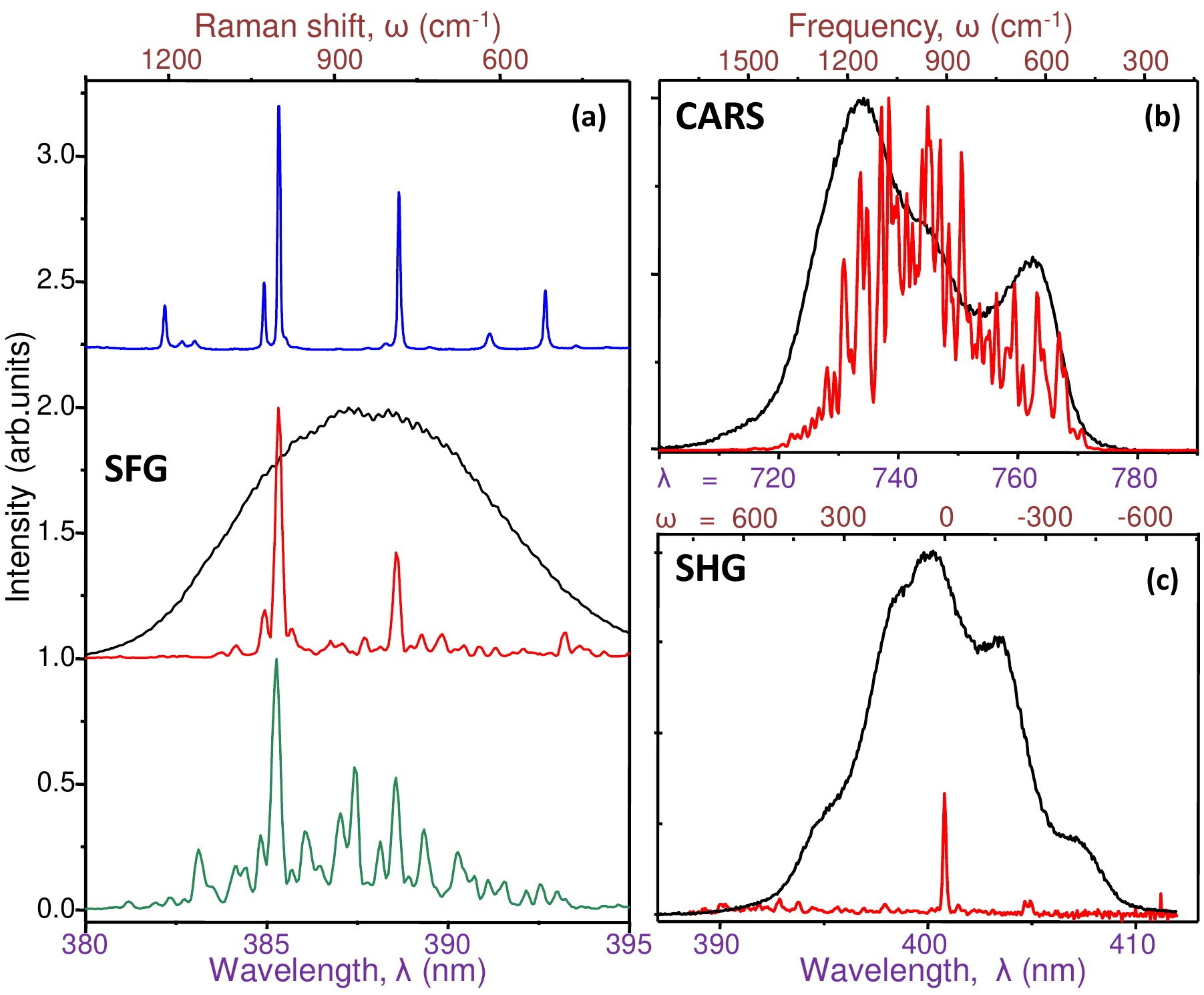}
\caption{(Color online) (a) Experimentally observed \textsc{cars}/probe \textsc{sfg} spectra (middle and bottom) in comparison to the spectrum of spontaneous Raman scattering (top). Black, green and red curves correspond to the case of unshaped, non-optimized shaped and optimized shaped probe pulse, respectively. (b) \textsc{cars} spectra for the shaped (red) and unshaped (black) probe pulses. (c) The spectra of the second harmonic of the shaped (red) and unshaped (black) probe pulses.}
  %\vskip -.1truein
  \label{FigResults}
\end{figure}
Experimental \textsc{sfg} spectra are presented in Fig.\ref{FigResults}(a). The spontaneous Raman spectrum of toluene, measured separately, is shown at the top of the plot for reference. Below it, the two signals with sharp peaks of much narrower width than the bandwidth of the probe pulse, attest to the high resolution of the proposed method. The position of the two peaks of the center curve match the frequencies of the two strong Raman modes of toluene around 1000 and 784 cm$^{-1}$, covered by the broad bandwidth of the pump/Stokes pulse pair. The latter is demonstrated by the observed \textsc{cars} signal, with and without phase shaping, in plot (b). The apparently random \textsc{cars} intensity distribution in the shaped case results from the presence of Raman modes and carries the information about their frequencies, which is revealed in the \textsc{cars}/probe \textsc{sfg} spectrum.

The \textsc{sfg} spectrum at the bottom of Fig.\ref{FigResults}(a) exhibits low signal-to-noise ratio as a result of the lack of the phase mask optimization. An optimized mask has resulted in higher quality spectrum shown in the middle of the plot. Note that the optimization has not been performed during the experiment in an adaptive feedback-controlled fashion, in which the observed spectral contrast would have been used as a fitness parameter. Instead, we have pre-calculated our binary phase masks for the anticipated resonance line width and the measured characteristics of the probe pulse (i.e. its spectral intensity and phase before shaping). The latter proves critical for the quality of the observed \textsc{sfg} signal. We therefore attribute the unequal performance of different masks to the uncertainty in characterizing the probe pulse, which will be improved in the future.

As expected, spectral narrowing of the \textsc{cars}/probe \textsc{sfg} signal is accompanied by the corresponding narrowing of the probe's second harmonic, shown in Fig.\ref{FigResults}(c). We have found that for our experimental parameters, better peak-to-background contrast in the \textsc{sfg} always corresponds to the better contrast of the \textsc{shg} peak. The reason for this is the insufficient resolution of our pulse shaper for resolving the line width of the observed Raman modes (10 cm$^{-1}$ vs 3 cm$^{-1}$, respectively). As a result, the optimization is almost insensitive to the width of the resonant lines and works equally well for both the \textsc{sfg} and \textsc{shg} narrowing. This however, may not be true for the case of higher resolution shaping.

%-----------------------------Conclusion-----------------------------------------------
The demonstrated approach presents an alternative to the existing methods of high-resolution Raman spectroscopy with ultrashort pulses, which typically rely on the delay scanning, interferometric stability, or post-processing of the acquired data. All-optical correlation analysis is performed separately from the resonant medium of interest, reducing the sensitivity of the method to non-resonant background, precise time delays, and other laser parameters. The ability to detect Raman peaks directly in the measured \textsc{sfg} spectrum may prove invaluable for the applications in microscopy. For instance, sensitive detection of the integral \textsc{sfg} signal, passed through an appropriate frequency filter, could quickly identify the presence of a certain Raman mode without the need of a spectrometer. Though this type of detection is similar to the traditional spontaneous Raman or multiplex \textsc{cars} spectroscopy, the proposed method utilizes the whole spectral bandwidth and high peak power of femtosecond pulses, thus suggesting the possibility for higher sensitivity or faster data retrieval. In the proof-of-principle work presented here, the speed was limited by the low efficiency of the sum-frequency generation. This can be improved by using thicker nonlinear crystals with optimized dispersion properties, and by increasing the power of a reference pulse. Pre-calculated phase masks may enable selective detection of resonances with the specified line width, while avoiding time consuming feedback-controlled optimization.

\begin{acknowledgements}
This work has been supported by the CFI, BCKDF and NSERC.
\end{acknowledgements}

%-----------Bibliography ---------------------------------
%\bibliography{Nascars2}

\begin{thebibliography}{28}
\expandafter\ifx\csname natexlab\endcsname\relax\def\natexlab#1{#1}\fi
\expandafter\ifx\csname bibnamefont\endcsname\relax
  \def\bibnamefont#1{#1}\fi
\expandafter\ifx\csname bibfnamefont\endcsname\relax
  \def\bibfnamefont#1{#1}\fi
\expandafter\ifx\csname citenamefont\endcsname\relax
  \def\citenamefont#1{#1}\fi
\expandafter\ifx\csname url\endcsname\relax
  \def\url#1{\texttt{#1}}\fi
\expandafter\ifx\csname urlprefix\endcsname\relax\def\urlprefix{URL }\fi
\providecommand{\bibinfo}[2]{#2}
\providecommand{\eprint}[2][]{\url{#2}}

\bibitem[{\citenamefont{Joo {\it et~al.}}(1991)\citenamefont{Joo, Dugan, and
  Albrecht}}]{Joo91}
\bibinfo{author}{\bibfnamefont{T.}~\bibnamefont{Joo}},
  \bibinfo{author}{\bibfnamefont{M.~A.} \bibnamefont{Dugan}}, \bibnamefont{and}
  \bibinfo{author}{\bibfnamefont{A.~C.} \bibnamefont{Albrecht}},
  \bibinfo{journal}{Chem. Phys. Lett.} \textbf{\bibinfo{volume}{177}},
  \bibinfo{pages}{4} (\bibinfo{year}{1991}).

\bibitem[{\citenamefont{Schmitt {\it et~al.}}(1997)}]{Schmitt97}
\bibinfo{author}{\bibfnamefont{M.}~\bibnamefont{Schmitt}} \bibnamefont{{\it
  et~al.}}, \bibinfo{journal}{Chem. Phys. Lett.}
  \textbf{\bibinfo{volume}{270}}, \bibinfo{pages}{9} (\bibinfo{year}{1997}).

\bibitem[{\citenamefont{Lang {\it et~al.}}(1999)\citenamefont{Lang, Kompa, and
  Motzkus}}]{Lang99}
\bibinfo{author}{\bibfnamefont{T.}~\bibnamefont{Lang}},
  \bibinfo{author}{\bibfnamefont{K.-L.} \bibnamefont{Kompa}}, \bibnamefont{and}
  \bibinfo{author}{\bibfnamefont{M.}~\bibnamefont{Motzkus}},
  \bibinfo{journal}{Chem. Phys. Lett.} \textbf{\bibinfo{volume}{310}},
  \bibinfo{pages}{65} (\bibinfo{year}{1999}).

\bibitem[{\citenamefont{Scully {\it et~al.}}(2002)}]{Scully02}
\bibinfo{author}{\bibfnamefont{M.~O.} \bibnamefont{Scully}} \bibnamefont{{\it
  et~al.}}, \bibinfo{journal}{PNAS} \textbf{\bibinfo{volume}{99}},
  \bibinfo{pages}{10994} (\bibinfo{year}{2002}).

\bibitem[{\citenamefont{Ooi {\it et~al.}}(2005)}]{Ooi05}
\bibinfo{author}{\bibfnamefont{C.~H.~R.} \bibnamefont{Ooi}} \bibnamefont{{\it
  et~al.}}, \bibinfo{journal}{Phys. Rev. A} \textbf{\bibinfo{volume}{72}},
  \bibinfo{pages}{023807} (\bibinfo{year}{2005}).

\bibitem[{\citenamefont{Zumbusch {\it et~al.}}(1999)\citenamefont{Zumbusch,
  Holtom, and Xie}}]{Zumbusch99}
\bibinfo{author}{\bibfnamefont{A.}~\bibnamefont{Zumbusch}},
  \bibinfo{author}{\bibfnamefont{G.~R.} \bibnamefont{Holtom}},
  \bibnamefont{and} \bibinfo{author}{\bibfnamefont{X.~S.} \bibnamefont{Xie}},
  \bibinfo{journal}{Phys. Rev. Lett.} \textbf{\bibinfo{volume}{82}},
  \bibinfo{pages}{4142} (\bibinfo{year}{1999}).

\bibitem[{\citenamefont{Potma {\it et~al.}}(2001)}]{Potma01}
\bibinfo{author}{\bibfnamefont{E.~O.} \bibnamefont{Potma}} \bibnamefont{{\it
  et~al.}}, \bibinfo{journal}{PNAS} \textbf{\bibinfo{volume}{98}},
  \bibinfo{pages}{1577} (\bibinfo{year}{2001}).

\bibitem[{\citenamefont{Cheng and Xie}(2004)}]{Cheng04}
\bibinfo{author}{\bibfnamefont{J.~X.} \bibnamefont{Cheng}} \bibnamefont{and}
  \bibinfo{author}{\bibfnamefont{X.~S.} \bibnamefont{Xie}},
  \bibinfo{journal}{J. Phys. Chem. B} \textbf{\bibinfo{volume}{108}},
  \bibinfo{pages}{827} (\bibinfo{year}{2004}).

\bibitem[{\citenamefont{Heid {\it et~al.}}(2001)}]{Heid01}
\bibinfo{author}{\bibfnamefont{M.}~\bibnamefont{Heid}} \bibnamefont{{\it
  et~al.}}, \bibinfo{journal}{Journal of Raman Spectroscopy}
  \textbf{\bibinfo{volume}{32}}, \bibinfo{pages}{771} (\bibinfo{year}{2001}).

\bibitem[{\citenamefont{Xu {\it et~al.}}(2007)}]{Xu07}
\bibinfo{author}{\bibfnamefont{X.~G.} \bibnamefont{Xu}} \bibnamefont{{\it
  et~al.}}, \bibinfo{journal}{J. Chem. Phys.} \textbf{\bibinfo{volume}{126}}
  (\bibinfo{year}{2007}).

\bibitem[{\citenamefont{Cheng {\it et~al.}}(2002)}]{Cheng02}
\bibinfo{author}{\bibfnamefont{J.~X.} \bibnamefont{Cheng}} \bibnamefont{{\it
  et~al.}}, \bibinfo{journal}{J. Phys. Chem. B} \textbf{\bibinfo{volume}{106}},
  \bibinfo{pages}{8493} (\bibinfo{year}{2002}).

\bibitem[{\citenamefont{Oron {\it et~al.}}(2003)\citenamefont{Oron, Dudovich,
  and Silberberg}}]{Oron03}
\bibinfo{author}{\bibfnamefont{D.}~\bibnamefont{Oron}},
  \bibinfo{author}{\bibfnamefont{N.}~\bibnamefont{Dudovich}}, \bibnamefont{and}
  \bibinfo{author}{\bibfnamefont{Y.}~\bibnamefont{Silberberg}},
  \bibinfo{journal}{Phys. Rev. Lett.} \textbf{\bibinfo{volume}{90}},
  \bibinfo{pages}{213902} (\bibinfo{year}{2003}).

\bibitem[{\citenamefont{Lim {\it et~al.}}(2005)\citenamefont{Lim, Caster, and
  Leone}}]{Lim05}
\bibinfo{author}{\bibfnamefont{S.-H.} \bibnamefont{Lim}},
  \bibinfo{author}{\bibfnamefont{A.~G.} \bibnamefont{Caster}},
  \bibnamefont{and} \bibinfo{author}{\bibfnamefont{S.~R.} \bibnamefont{Leone}},
  \bibinfo{journal}{Phys. Rev. A} \textbf{\bibinfo{volume}{72}},
  \bibinfo{pages}{041803} (\bibinfo{year}{2005}).

\bibitem[{\citenamefont{Weiner}(2000)}]{Weiner00}
\bibinfo{author}{\bibfnamefont{A.~M.} \bibnamefont{Weiner}},
  \bibinfo{journal}{Rev. Sci. Instrum.} \textbf{\bibinfo{volume}{71}},
  \bibinfo{pages}{1929} (\bibinfo{year}{2000}).

\bibitem[{\citenamefont{Zheng and Weiner}(2001)}]{Zheng01}
\bibinfo{author}{\bibfnamefont{Z.}~\bibnamefont{Zheng}} \bibnamefont{and}
  \bibinfo{author}{\bibfnamefont{A.~M.} \bibnamefont{Weiner}},
  \bibinfo{journal}{Chem. Phys.} \textbf{\bibinfo{volume}{267}},
  \bibinfo{pages}{161} (\bibinfo{year}{2001}).

\bibitem[{\citenamefont{Meshulach and Silberberg}(1998)}]{Meshulach98}
\bibinfo{author}{\bibfnamefont{D.}~\bibnamefont{Meshulach}} \bibnamefont{and}
  \bibinfo{author}{\bibfnamefont{Y.}~\bibnamefont{Silberberg}},
  \bibinfo{journal}{Nature} \textbf{\bibinfo{volume}{396}},
  \bibinfo{pages}{239} (\bibinfo{year}{1998}).

\bibitem[{\citenamefont{Pastirk {\it et~al.}}(2003)}]{Pastirk03}
\bibinfo{author}{\bibfnamefont{I.}~\bibnamefont{Pastirk}} \bibnamefont{{\it
  et~al.}}, \bibinfo{journal}{Opt. Express} \textbf{\bibinfo{volume}{11}},
  \bibinfo{pages}{1695} (\bibinfo{year}{2003}).

\bibitem[{\citenamefont{Lozovoy {\it et~al.}}(2003)}]{Lozovoy03}
\bibinfo{author}{\bibfnamefont{V.~V.} \bibnamefont{Lozovoy}} \bibnamefont{{\it
  et~al.}}, \bibinfo{journal}{J. Chem. Phys.} \textbf{\bibinfo{volume}{118}},
  \bibinfo{pages}{3187} (\bibinfo{year}{2003}).

\bibitem[{\citenamefont{Dudovich {\it et~al.}}(2002)\citenamefont{Dudovich,
  Oron, and Silberberg}}]{Dudovich02}
\bibinfo{author}{\bibfnamefont{N.}~\bibnamefont{Dudovich}},
  \bibinfo{author}{\bibfnamefont{D.}~\bibnamefont{Oron}}, \bibnamefont{and}
  \bibinfo{author}{\bibfnamefont{Y.}~\bibnamefont{Silberberg}},
  \bibinfo{journal}{Nature} \textbf{\bibinfo{volume}{418}},
  \bibinfo{pages}{512} (\bibinfo{year}{2002}).

\bibitem[{\citenamefont{Oron {\it et~al.}}(2002)}]{Oron02}
\bibinfo{author}{\bibfnamefont{D.}~\bibnamefont{Oron}} \bibnamefont{{\it
  et~al.}}, \bibinfo{journal}{Phys. Rev. Lett.} \textbf{\bibinfo{volume}{88}},
  \bibinfo{pages}{063004} (\bibinfo{year}{2002}).

\bibitem[{\citenamefont{Zheltikov}(2000)}]{Zheltikov00}
\bibinfo{author}{\bibfnamefont{A.~M.} \bibnamefont{Zheltikov}},
  \bibinfo{journal}{Journal of Raman Spectroscopy}
  \textbf{\bibinfo{volume}{31}}, \bibinfo{pages}{653} (\bibinfo{year}{2000}).

\bibitem[{\citenamefont{Nibbering {\it et~al.}}(1992)\citenamefont{Nibbering,
  Wiersma, and Duppen}}]{Nibbering92}
\bibinfo{author}{\bibfnamefont{E.~T.~J.} \bibnamefont{Nibbering}},
  \bibinfo{author}{\bibfnamefont{D.~A.} \bibnamefont{Wiersma}},
  \bibnamefont{and} \bibinfo{author}{\bibfnamefont{K.}~\bibnamefont{Duppen}},
  \bibinfo{journal}{Phys. Rev. Lett.} \textbf{\bibinfo{volume}{68}},
  \bibinfo{pages}{514} (\bibinfo{year}{1992}).

\bibitem[{\citenamefont{Xu {\it et~al.}}(2008)}]{Xu08}
\bibinfo{author}{\bibfnamefont{X.~G.} \bibnamefont{Xu}} \bibnamefont{{\it
  et~al.}}, \bibinfo{journal}{Nat Phys} \textbf{\bibinfo{volume}{4}},
  \bibinfo{pages}{125} (\bibinfo{year}{2008}).

\bibitem[{\citenamefont{Diels and Rudolph}(2006)}]{DielsRudolphBook}
\bibinfo{author}{\bibfnamefont{J.-C.} \bibnamefont{Diels}} \bibnamefont{and}
  \bibinfo{author}{\bibfnamefont{W.}~\bibnamefont{Rudolph}},
  \emph{\bibinfo{title}{Ultrashort Laser Pulse Phenomena}}
  (\bibinfo{publisher}{Elsevier}, \bibinfo{year}{2006}), \bibinfo{edition}{2nd}
  ed.

\bibitem[{\citenamefont{Oron {\it et~al.}}(2004)\citenamefont{Oron, Dudovich,
  and Silberberg}}]{Oron04}
\bibinfo{author}{\bibfnamefont{D.}~\bibnamefont{Oron}},
  \bibinfo{author}{\bibfnamefont{N.}~\bibnamefont{Dudovich}}, \bibnamefont{and}
  \bibinfo{author}{\bibfnamefont{Y.}~\bibnamefont{Silberberg}},
  \bibinfo{journal}{Phys. Rev. A} \textbf{\bibinfo{volume}{70}},
  \bibinfo{pages}{023415} (\bibinfo{year}{2004}).

\bibitem[{\citenamefont{Dayan {\it et~al.}}(2004)}]{Dayan04}
\bibinfo{author}{\bibfnamefont{B.}~\bibnamefont{Dayan}} \bibnamefont{{\it
  et~al.}}, \bibinfo{journal}{Phys. Rev. Lett.} \textbf{\bibinfo{volume}{93}},
  \bibinfo{pages}{023005} (\bibinfo{year}{2004}).

\bibitem[{\citenamefont{Comstock {\it et~al.}}(2004)}]{Comstock04}
\bibinfo{author}{\bibfnamefont{M.}~\bibnamefont{Comstock}} \bibnamefont{{\it
  et~al.}}, \bibinfo{journal}{Opt. Express} \textbf{\bibinfo{volume}{12}},
  \bibinfo{pages}{1061} (\bibinfo{year}{2004}).

\bibitem[{\citenamefont{Zheng and Weiner}(2000)}]{Zheng00}
\bibinfo{author}{\bibfnamefont{Z.}~\bibnamefont{Zheng}} \bibnamefont{and}
  \bibinfo{author}{\bibfnamefont{A.~M.} \bibnamefont{Weiner}},
  \bibinfo{journal}{Opt. Lett.} \textbf{\bibinfo{volume}{25}},
  \bibinfo{pages}{984} (\bibinfo{year}{2000}).

\end{thebibliography}

\end{document}